# Some comments on Laplacian gauge fixing

Pierre van Baal [a] [*]

[a]Instituut-Lorentz for Theoretical Physics,
University of Leiden, PO Box 9506,
NL-2300 RA Leiden, The Netherlands.

Laplacian gauge fixing was introduced [1] to find a unique representative of the gauge orbit, which on the lattice could be implemented by a "finite" algorithm. What was still lacking was a perturbative formulation of this gauge, which will be presented here. However, renormalizability is still to be demonstrated. For torodial and spherical geometries a detailed comparison with the Landau (or Coulomb) gauge will be made.

In perturbation theory, gauge fixing on the lattice is not much different from the continuum, apart from some technical complications. The advantage of the lattice formulation is that one can go beyond perturbation theory and take the contribution of large fields into account, where the issue of finding a unique representative on the gauge orbit becomes essential. Frequently gauge fixing on the lattice is used to define an operator that is difficult to define in a gauge invariant way or for which the gauge invariant version couples to the required physical state with too small an amplitude. The gauge field configurations are still generated with the standard Wilson measure, meaning that formally the Monte Carlo average also includes an average over the gauge orbit with the appropriate weight factors, thereby projecting the gauge fixed operator on a gauge invariant one. If the gauge fixing is, however, improperly implemented one is likely to lose control over what the precise nature of this projected gauge invariant operator will be.

Until recently, the most popular gauge fixing on the lattice was to implement Landau or Coulomb gauge fixing by finding the maximum of $\sum_{x,\mu} \text{Tr}(U_\mu(x))$ as the unique representative on the gauge orbit. This is, however, similar to a spin-glass problem, with many local maxima, and there does not exist an algorithm that allows one to find the absolute maximum [2]. In the continuum the equivalent formulation would be to minimize the $L_2$ norm of the gauge field, $||A||^2 \equiv \int \text{Tr}(A_\mu(x) A_\mu^\dagger(x))$, along the gauge orbit. An alternative was formulated by Vink and Wiese [1] by considering the eigenfunctions of the covariant Laplacian with the lowest eigenvalues. To keep things transparent we will in the following work in the continuum for the gauge group SU(2). For the lattice formulation and for general gauge groups see refs. [1,2].

## 1. The Laplace gauge

Consider $(-)$ the covariant Laplacian $-D_\mu^2(A)$, where $D_\mu(A) = \partial_\mu + A_\mu$. It is a positive operator with a two-fold degenerate spectrum due to charge conjugation symmetry. Parametrizing the gauge field in terms of the Pauli matrices $\sigma_a$ as $A_\mu = iA_\mu^a \sigma_a/2$ one has $A_\mu^* = (i\sigma_2) A_\mu (i\sigma_2)^\dagger$. The eigenfunction $h = (h_1, h_2)^t$ can be combined with $(i\sigma_2)^\dagger h^* = (-h_2^*, h_1^*)^t$ (having the same eigenvalue) into a quaternion $q \equiv q_0 1_2 + i\vec{q}\cdot\vec{\sigma}$, with $h_1 = q_0 + iq_3$ and $h_2 = i(q_1 + iq_2)$. The vector potential is a so-called imaginary quaternion (with $q_0 = 0$), whereas a gauge function $g(x) \in \text{SU}(2)$ is a unit quaternions ($|q|^2 \equiv \sum_{k=0}^3 q_k^2 = 1$). The covariant Laplacian acts on $q$ by matrix multiplication. Under a gauge transformation one finds

$$A_\mu \to g^{-1} A_\mu g + g^{-1} \partial_\mu g \quad , \quad q \to g^{-1} \cdot q \quad . \quad (1)$$

The quaternion $q$ can be written as $q = g|q|$, which is unique if there are no further degeneracies and is well-defined if $|q(x)| \neq 0$ for all $x$. The Laplace gauge is defined by the following proce-

---

dure [1]. Pick any $A$ on the gauge orbit. Find its *lowest* eigenvalue with eigenfunction $q$, determine $g(x) \equiv q(x)/|q(x)|$ and transform $A_\mu$ to $g^{-1}D_\mu(A)g$. To derive the perturbative formulation of the Laplace gauge, we observe that the local gauge condition clearly amounts to the lowest eigenfunction being of the form $q(x) = |q(x)|1_2$. Calling $\lambda_0$ the lowest eigenvalue, we can re-write the covariant Laplace equation in terms of a real quaternion (i.e. a scalar) and an imaginary quaternion (i.e. an SU(2) Lie-algebra element)

$$(D_\mu^2 + \lambda_0)q = (\partial_\mu^2 + A_\mu^2 + \lambda_0)|q| + |q|^{-1}\partial_\mu(|q|^2 A_\mu), \quad (2)$$

where $A$ has been brought to the Laplace gauge. We have therefore separately that $\partial_\mu(|q|^2 A_\mu) = 0$ and $(\partial_\mu^2 + A_\mu^2 + \lambda_0)|q| = 0$. The first relation is very similar to the Landau (or Coulomb) gauge, but $|q|$ depends on the vector potential $A$. However, since $|q|$ is gauge invariant, it depends only on the gauge orbit as a whole, and not on the position of $A$ along the orbit. This therefore establishes the perturbative formulation of the Laplace gauge. Introducing ghosts $\bar c$ and $c$ would lead to adding to the action the ghost part

$$S_{\rm gh} = \int d_n x \; s\bar c \left(b - \partial_\mu(|q_A|^2 A^\mu)\right) \qquad (3)$$

$$= \int \left(b^2 - b\partial_\mu(|q_A|^2 A^\mu) - |q_A|^2 \partial_\mu \bar c D^\mu c\right) d_n x,$$

where $s$ is the usual BRST generator, $sA_\mu = D_\mu c$, $sc = -\frac{1}{2}[c,c]$, $s\bar c = b$ and $sb = 0$. The presence of $|q_A|$ introduces non-localities and renormalizability is far from obvious. It might be that the Ward identities derived from the BRST invariance will nevertheless show that the quantum theory is well-defined.

The condition $\partial_\mu(|q_A|^2 A^\mu) = 0$ can, as for the Landau gauge, be formulated in terms of the absolute minimum of a functional defined by $||A||_q^2 \equiv \int |q(x)|^2 {\rm Tr}(A_\mu^\dagger(x)A_\mu(x))$. If we assume $A$ is in the Laplace gauge and $q \equiv q_A$ one finds

$$||{}^g A||_q^2 - ||A||_q^2 = 2 \int |q|^2 \left(|g^\dagger D_\mu g|^2 - |A_\mu|^2\right)$$

$$= 2 \int |D_\mu(|q|g)|^2 - \lambda_0 ||q|g|^2 \geq 0, \qquad (4)$$

where we used eq. (2) and the fact that $|g(x)| = 1$. The two norms are of course equal if $g(x)$ is constant. Like in the Landau gauge, $\partial_\mu(|q_A|^2 A^\mu) = 0$ fixes the gauge only up to a constant gauge transformation. The only other case where the two norms can coincide is when the covariant Laplacian has an accidental degeneracy *and* its two eigenfunctions are related by a (non-constant) gauge transformation.

## 2. Examples

The simplest set of examples is obtained by taking $A_\mu^2(x)$ to be constant (since it is a scalar, this is true for any choice of coordinates). It follows from eq. (4) and the fact that $|q| = 1$, that in this case the Laplace gauge coincides with the Coulomb or Landau gauge. For a torus with sides of length $L = 1$, the constant abelian modes $A_\mu(x) = iC_\mu\sigma_3/2$ were studied in this context before [3]. For the gauge group SO(3) the fundamental domain, being the set of absolute minima of the norm functional, coincides with a torus of length $2\pi$ centred at $A = 0$. Its boundary is exactly where the covariant Laplacian has a degenerate eigenvalue, since the eigenfunctions are given by $h_k(x) = \exp(\pi i x \cdot k \sigma_3)h_0$, with the eigenvalues $\lambda_k = \frac{1}{4}(2\pi k_\mu - C_\mu)^2$, where $k \in \mathbb{Z}^n$ and $h_0$ is an arbitrary constant spinor (representing the generic two-fold degeneracy). Moving out from the origin, $A = 0$, the first crossing of the lowest two eigenvalues $\lambda_k$ is easily seen to occur at the convex hull of the planes specified by $k_\mu = \pm\pi$ for each $\mu$. Of course it is still required to divide out the constant gauge transformations (in this case mapping $C$ to $-C$) to obtain the true fundamental domain. Opposite points on its boundary are identified by the anti-periodic gauge transformations $g^{\pm\mu}(x) = \exp(\pm\pi i x_\mu \sigma_3)$, which furthermore relate the eigenfunctions for the two lowest eigenvalues that cross at the boundary of the fundamental domain.

For all cases where $A_\mu^2(x)$ is constant, there is a one to one relation between the fundamental domains in both gauges and for both, points at the boundary are identified by suitable gauge transformations [2,3]. As an other example, consider gauge fields on $S^3$ [4]. One introduces a framing (for the embedding $n_\mu \equiv x_\mu/|x|$ in $\mathbb{R}^4$) in terms of the 't Hooft self-dual and anti-selfdual tensors, $e_\mu^a \equiv \eta_{\mu\nu}^a n_\nu$ and $\bar e_\mu^a \equiv \bar\eta_{\mu\nu}^a n_\mu$. The dreibeins

$e_\mu^a$ and $\bar{e}_\mu^a$ have opposite orientations and are related to each other by a gauge transformation, $g = n_0 + i\vec{n}\cdot\vec{\sigma}$, with winding number one, with [4] $g^\dagger e_\mu^a \sigma_a g = -\bar{e}_\mu^a \sigma_a$. Up to a gauge and a rotation, constant $A_\mu^a(x)$ is defined by $A_\mu^a = c_i^a e_\mu^i$, with $c_i^a = y_i \delta_{ia}$. In terms of $\vec{y}$ the fundamental domain forms a tetrahedron centred at the origin, whose edges coincide with the Gribov horizon.

To explicitly demonstrate that there is *in general* no one to one relation between the two gauges, we now consider on $S^3$ the gauge field $A_\mu^a = -u e_\mu^a - v \bar{e}_\mu^a$, which satisfies $\partial_\mu A^\mu = 0$, but is no longer constant for all $u$ and $v$. Introducing the commuting angular momentum operators $\vec{T} = -\frac{1}{2}\vec{\sigma}$, $\vec{L}_1 = \frac{1}{2i}\vec{e}_\mu \partial_\mu$ and $\vec{L}_2 = \frac{1}{2i}\vec{\bar{e}}_\mu \partial_\mu$, one finds $-D_\mu^2 = 4\vec{L}_1^2 + 4(u\vec{L}_1 + v\vec{L}_2)\cdot\vec{T} + \frac{3}{4}(u^2+v^2) - \frac{uv}{2}(4n_0^2 - 1)$. It commutes with $\vec{J} \equiv \vec{L}_1 + \vec{L}_2 + \vec{T}$ and can be classified by $j^\pi$, where $\pi = (-1)^{2\ell}$ is the parity, with $\vec{L}_1^2 = \vec{L}_2^2 = \ell(\ell+1)$, $\ell = 0, \frac{1}{2}, 1, \cdots$. Since $T$ carries spin one-half, $j$ is always half-integer. The spectrum of the covariant Laplacian is independent of $j_z$, where charge conjugation relates $j_z$ to $-j_z$. It is easy to write down a complete basis (introducing the short-hand notation $j_\pm = j \pm \frac{1}{2} - \frac{1}{2}$)

$$|\ell,j,j_z\rangle_\pm = \sum (-1)^{m_3-1+j_\pm} \sqrt{2j_\pm(2j+1)} |\tfrac{1}{2},m_3\rangle \times$$
$$\begin{pmatrix} \frac{1}{2}+j_\pm & \frac{1}{2} & j \\ m_1+m_2 & m_3 & -j_z \end{pmatrix} \begin{pmatrix} \ell & \ell & \frac{1}{2}+j_\pm \\ m_1 & m_2 & -m_1-m_2 \end{pmatrix} |\ell,m_1,m_2\rangle,$$

where $|\ell,m_1,m_2\rangle \equiv S_{2\ell}^{m_1,m_2}(\xi = n_1 + in_2, \zeta = n_3 + in_0)$ as defined in ref. [5], and $|\frac{1}{2},\pm\frac{1}{2}\rangle$ are the spin-up and down spinors. Matrix elements of the covariant Laplacian can be easily computed algebraically in this basis. For $(u^2+v^2) < 30$ and all relevant values of $j^\pi$ (i.e. $j < 3$), the lowest eigenvalues can be computed using less than twelve basis functions in each sector with the upper *and* lower bounds coinciding with an accuracy of better than 1 part in $10^6$. The figure labels regions by the values of $j^\pi$ for the lowest eigenvalue of the covariant Laplacian. It can be rigorously shown that $|q(x)|$ vanishes for $u = v > 0.77703$ at the two poles and for $-2.28491 < u = v < -2.24641$ in addition along the equator of $S^3$. This can also be seen from the fact that the eigenfunctions $q(x)$ depend smoothly on $u$ and $v$, but that the relevant gauge functions $q(x)/|q(x)|$ (for $j^\pi = 1/2^-$ or $3/2^-$) flip their winding number (in absolute value equal to one), when crossing the line $u = v$. The only way they can change their winding number, is by having somewhere (at least two) zero's. Finally, it is conjectured that in general inside $\tilde{\Lambda} \equiv \{A | \partial_\mu A^\mu = 0 \text{ and } -D^\mu \partial_\mu \geq 0\}$ (indicated in fig. 1 by the dashed curve, which coincides with the Gribov horizon along $u+v = -3$) no ground-state level crossings occur. Note that the fundamental domain $\Lambda$ (thin curves bounded by the Gribov horizon) contains [4] $\tilde{\Lambda}$, *but has crossings*.

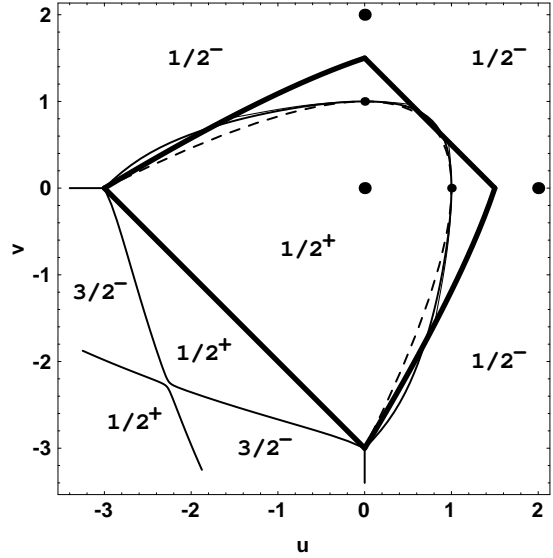

Figure 1. Crossings of the lowest eigenvalues of $-D_\mu^2(A)$, the Gribov horizon (fat section), $\partial\tilde{\Lambda}$ (dashed curves), $\partial\Lambda$ (thin lines), classical vacua (large dots) and the sphalerons (smaller dots).

Discussions with Jeroen Vink at various occasions are gratefully acknowledged. I also thank the Aspen Center for Physics, where some of this work was done, for its hospitality in July 1994.